\documentclass[aps,prl,floatfix,amsmath,amssymb,twocolumn,footinbib]{revtex4-1}
\usepackage{epsfig}
\usepackage{natbib}
\usepackage{color}
\newcommand{\diff}{\mathrm{d}}
\begin{document}

\title{General continuum approach for dissipative systems of
  repulsive particles} \author{C\'esar M. Vieira, Humberto A. Carmona,
  Jos\'e S. Andrade Jr., and Andr\'e A. Moreira}
\affiliation{Departamento de F\'isica, Universidade Federal do
  Cear\'a, 60451-970 Fortaleza, Brazil}

\begin{abstract} 
  We propose a general coarse-graining method to derive a continuity
  equation that describes any dissipative system of repulsive
  particles interacting through short-ranged potentials. In our
  approach, the effect of particle-particle correlations is
  incorporated to the overall balance of energy, and a non-linear
  diffusion equation is obtained to represent the overdamped
  dynamics. In particular, when the repulsive interaction potential is
  a short-ranged power-law, our approach reveals a distinctive
  correspondence between particle-particle energy and the generalized
  thermostatistics of Tsallis for any non-positive value of the
  entropic index $q$. Our methodology can also be applied to
  microscopic models of superconducting vortices and complex plasma,
  where particle-particle correlations are pronounced at low
  concentrations. The resulting continuum descriptions provide
  elucidating and useful insights on the microdynamical behavior of
  these physical systems. The consistency of our approach is
  demonstrated by comparison with molecular dynamics simulations.
\end{abstract}\maketitle

Dissipative systems of repulsive particles can be representative of
many physical phenomena in nature. For instance, type-II
superconductors can be populated by vortices of super-currents that
pierce the system in the direction of the applied magnetic
field~\cite{gennes}. These vortices can be considered as single
particles, and will dissipate energy as long as they
move~\cite{ref8,ref9_1,*ref9_2,ref9_3,ref10_1,*ref10_2,*ref11_1,*ref11_2,
  ref12,*ref13_1,*ref13_2}. Another notable example is complex plasma
\cite{shukla,hiroo}, where charged micro-particles are immersed in an
ionized gas.  The free charges in the plasma screen the electric
repulsion between particles, and electro-mechanical couplings between
particles and media act as a drag force to the movement
\cite{shukla}. Colloidal systems are also well described with such
approach \cite{diego1,*diego2,munarin}.

A general equation of motion for systems of interacting particles in a
dissipative media can be written as,
\begin{equation}\label{eq:amort}
  m_i\frac{\diff{\mathbf{v}_i}}{\diff{t}}=
  \sum_j\mathbf{F}_{ij}+
  \sum_e\mathbf{F}_e 
  -\gamma\mathbf{v}_i
  +\sqrt{2\gamma k_{B}T}\boldsymbol{\eta}_i(t),
\end{equation}
where $\mathbf{v}_i$ is the velocity of a single particle $i$.  The
first summation in Eq.~(\ref{eq:amort}) goes over the forces due to
other particles.  We refer to the effect of the other particles in the
system as the internal force $\mathbf{F}_{int}=\sum_j\mathbf{F}_{ij}$.
The second summation in Eq.~(\ref{eq:amort}) accounts for the action
of external fields on the particle, for instance, the electric and
gravitational fields in the case of complex plasma, or applied
electric currents in the case of superconducting vortices.  We use
$\mathbf{F}_{ext}(\mathbf{r}_i) = \sum_e\mathbf{F}_{e}$ to represent
the external forces acting on the particle.  The term
$-\gamma\mathbf{v}_i$ describes the dissipative force, and
$\sqrt{2\gamma{k}_BT}\boldsymbol{\eta}_i(t)$ represents the thermal
noise.  Often in such systems inertial effects and thermal noise can
be neglected, when compared to the other terms.  In these situations,
the system is said to obey an overdamped dynamics, where the velocity
of a particle is proportional to the resultant force acting on
it,\begin{equation}\gamma\mathbf{v}_i=\mathbf{F}_{i},\end{equation}
where $\mathbf{F}_i \equiv \mathbf{F}_{ext}+\mathbf{F}_{int}$.

Solving the equations of motion for a macroscopic system can be
unpractical. One possible approach is to describe the system by a
``coarse-grained'' continuous function $\rho(\mathbf{r},t)$ such that
$\rho(\mathbf{r},t)\diff{v}$ is the number of particles in the volume
$\diff{v}$, at any time $t$ and position $\mathbf{r}$. Since the
number of particles is conserved, continuity holds,
$\partial\rho(\mathbf{r},t)/\partial{t}=-\nabla\cdot\mathbf{J}(\mathbf{r},t)$,
where
$\mathbf{J}(\mathbf{r},t)\equiv\rho(\mathbf{r},t)\mathbf{v}(\mathbf{r})$,
with the field $\mathbf{v}(\mathbf{r})$ giving the velocity of the
particles in $\diff{v}$. The velocity field should be proportional to
the resultant force field
$\mathbf{f}_1=\mathbf{f}_{ext}+\mathbf{f}_{int}$ acting on each
particle near $\mathbf{r}$.  While the external force may depend
explicitly on the position,
$\mathbf{f}_{ext}=\mathbf{F}_{ext}(\mathbf{r})$, in order to determine
the force due to particle-particle interactions $\mathbf{f}_{int}$, it
is necessary to know the particle concentration profile.  Previous
efforts in this direction~\cite{zapperi2001flux} have considered that
the internal force should be proportional to the concentration
gradient, $\mathbf{f}_{int}=-a\nabla\rho$. Here we show that this
expression remains valid only if particle-particle correlations are
unimportant. We then generalize the approach developed
in~\cite{zapperi2001flux} by disclosing an analytical formalism to
account for these correlations in terms of a continuum model, with the
only restriction being that the interaction potential decreases fast
enough to be considered short-range. We demonstrate the usefulness of
our approach by comparing its predictions with the results from
numerical simulations.

To devise our continuum approach, we note that the potential energy of
a particle $U_1$ should be a function of the local concentration and
its derivatives.  Under conditions where the interaction has a
finite range of action, and $\nabla\rho(\mathbf{r})$ varies slowly
enough to be considered constant within this interaction range, it is
reasonable to disregard higher derivatives and use a first order
expansion to describe the surrounding concentration,
$\rho(\mathbf{r})=\rho_0+\mathbf{r}\cdot\nabla\rho_0$.  Moreover, for
repulsive interactions, the state of minimum energy is always
homogeneous, therefore any contribution of $\nabla\rho$ to the energy
$U_1$ should be in second order.  We conclude that, in a first order
approximation, the potential energy of a single particle is a function
of the local concentration only, $U_1{\equiv}U_1(\rho)$. 

In our continuum description the total potential energy is
\begin{equation}\label{eq:ut}
    U_T=\int\rho(\mathbf{r},t)U_1(\rho(\mathbf{r},t))\diff{v}.
\end{equation}
Considering that the single particle potential energy $U_1$ depends
implicitly on time through $\rho(\mathbf{r},t)$, the time variation of
the total potential energy is 
\begin{equation}\label{eq:ut1}
    \frac{\diff{U_T}}{\diff{t}}=
    \int\frac{\partial{\rho}}{\partial{t}}\left(U_1+\rho\frac{\diff{U_1}}{\diff\rho}\right)\diff{v}.
\end{equation}     
Let us define $W(\mathbf{r})$ as the term between parenthesis in
Eq.~(\ref{eq:ut1}). The continuity equation together with the identity
$W\nabla\cdot\mathbf{J}=\nabla\cdot(\mathbf{J}W)-\mathbf{J}\cdot\nabla{W}$
split Eq.~(\ref{eq:ut1}) into two integrals.  From Gauss theorem, as
long as there is no current $\mathbf{J}$ entering the border of the
system, the integral of $\nabla\cdot(\mathbf{J}W)$ vanishes, resulting
in
\begin{equation}\label{eq:ut2} 
    \frac{\diff{U_T}}{\diff{t}}=\int\mathbf{J}\cdot
    \nabla\left(U_1+\rho\frac{\diff{U_1}}{\diff\rho}\right)\diff{v}.
\end{equation} 
On the other hand, the variation of the potential energy is given by
the dissipated power,
$\diff{U}_T/\diff{t}=-\sum_i\mathbf{F}_i\cdot\mathbf{v}_i$, which in
the continuum description can be expressed as
\begin{equation}\label{eq:ut3}
        \frac{\diff{U_T}}{\diff{t}}=-\int\mathbf{J}\cdot\mathbf{f}_1~\diff{v},
\end{equation}
where we use $\mathbf{J}=\rho\mathbf{v}$.  Comparing
Eqs.~(\ref{eq:ut2}) and (\ref{eq:ut3}) we find
\begin{equation}\label{f1}
    \mathbf{f}_1=-\nabla\left(U_1+\rho\frac{\diff{U_1}}{\diff\rho}\right).
\end{equation} 
Considering $U_1=U_{ext}(\mathbf{r})+U_{int}(\rho)$, we obtain
$\mathbf{f}_{1}=\mathbf{f}_{ext}+\mathbf{f}_{int}$, with
$\mathbf{f}_{ext}=-\nabla{U_{ext}}$, and 
\begin{equation}\label{eq:a_definition}
    \mathbf{f}_{int}=-\nabla U_{int}= -a(\rho)\nabla\rho,
\end{equation}
with
\begin{equation}\label{novoa}
    a(\rho)=2\frac{\diff{U_{int}}}{\diff\rho}+\rho\frac{\diff^2{U_{int}}}{\diff\rho^2}.
\end{equation} 
Therefore, the function $a(\rho)$, and consequently the force
$\mathbf{f}_{int}$, are determined from the dependence of the particle-particle
potential energy $U_{int}$ on the concentration $\rho$.  Including this in the
continuity equation, we obtain
\begin{equation}\label{eq:continuity} 
    \gamma\frac{\partial \rho}{\partial t}=
    -\nabla\cdot\left[ \rho\left(\mathbf{f}_{ext} - a(\rho)\nabla\rho\right)\right], 
\end{equation} 
where the non-linear term $a(\rho)\rho\nabla\rho=-\rho\mathbf{f}_{int}$
accounts for the contribution of the inter-particle forces to the local current.

A simple approximation to the potential energy $U_{int}$ is obtained
by disregarding particle-particle correlations and assuming that the
concentration $\rho$ is a constant within the range of the potential,
leading to \begin{equation}\label{eq:uint_omega}
  U_{int}=\frac{\rho\Omega_D}{2}\int_0^\infty{V(r)r^{D-1}\diff{r}},\end{equation}
where $V(r)$ is the repulsive radial potential between particles
separated by a distance $r$, and $D$ is the dimensionality of the
system. In two dimensions $\Omega_2=2\pi$ is the angle of a
circumference, while in three dimensions $\Omega_3=4\pi$ is the solid
angle of a sphere. Substituting Eq.~(\ref{eq:uint_omega}) for
$U_{int}$ in Eq.~(\ref{novoa}), we conclude that $a$ is a constant,
and the internal force is proportional to the local gradient as,
\begin{equation}\label{eq:a_zapperi}
  \mathbf{f}_{int}=-\nabla\rho~\Omega_D\int_0^\infty{V(r)r^{D-1}\diff{r}}.
\end{equation}
For the two-dimensional case, Eq.~(\ref{eq:a_zapperi}) is consistent
with the form proposed
in~\cite{zapperi2001flux,prb_moreira,*petrucio}. As already mentioned,
this approach disregards particle-particle correlations that could be
relevant, and, therefore, it is a good approximation only under
certain conditions. 

As a matter of fact, dissipative systems of repulsive particles tend
to form structural lattices with at least local order.  The effect of
the local correlations is specially important for interactions in the
fashion of a power-law,
$V_\lambda(r)=\varepsilon\left(r/\sigma\right)^{-\lambda}$.  If the
exponent $\lambda$ is large enough, the force is short-ranged, but the
integral of Eq.~(\ref{eq:a_zapperi}) diverges at $r\rightarrow{0}$.
It is not likely, however, that two repulsive particles in an
dissipative medium will ever collide, therefore this divergence is not
physical.  A possible way to deal with correlations would be to consider an exclusion region of finite radius $r_{o}$ around each particle, which becomes smaller as the concentration grows, $r_{o}=\alpha\rho^{-\frac{1}{D}}$. In this way, the integral (\ref{eq:a_zapperi}) can be written as,
\begin{equation}\label{eq:a_infer}
  \mathbf{f}_{int}=-\nabla\rho~\Omega_D\int_{\alpha\rho^{-\frac{1}{D}}}^\infty{V(r)r^{D-1}\diff{r}},
\end{equation}
where the force $\mathbf{f}_{int}$ is still proportional to the gradient,
but now with $a\equiv{a(\rho)}$. For the case of a power-law
potential, $a(\rho)$ becomes finite and is given by
\begin{equation}\label{eq:a_errado}  
  a_\lambda(\rho)=\Omega_D\frac{\varepsilon\sigma^\lambda \alpha^{D-\lambda}}{\lambda-D}\rho^{\frac{\lambda}{D}-1},
\end{equation}
with the condition $\lambda>D$. Although this is just a qualitative
correction, it clearly shows that particle-particle correlations
affect the resultant force. Moreover, it also indicates that these
effects may be accounted for in a continuum model through replacement
of the constant $a$ by the function $a(\rho)$. In what follows we
propose a general way to obtain an estimate of the function
$a(\rho)$.

Due to the dissipative medium, the particles will form configurations
of low potential energy.  The least-energy state, or ground state,
depends on the form of the interaction, and on the particle
concentration. However, for repulsive interactions, this ground state
should be homogeneous, $\nabla\rho=\mathbf{0}$. Also, often the ground
state is a regular lattice. We propose that the structure formed by
the particles in the ground state could be used to calculate the
single particle energy,
\begin{equation}\label{eq:Uint}
U_{int}(\rho)=\frac{1}{2}\sum_v{V(r_v)},
\end{equation} 
where $V(r)$ is the pair interaction potential, and the sum is over
the vertices $v$ of the homogeneous lattice. The dependence on $\rho$
is implicit in the positions $r_v$. To compute the variation
$\diff{U_{int}}/\diff\rho$ it is useful to consider that the lattice
parameter of the homogeneous lattice should depend on the
concentration
$\ell\sim\rho^{-1/D}\Rightarrow(\diff\ell/\diff\rho)=-\ell/D\rho$. Note
also that the positions of the vertices are proportional to $\ell$,
that is, $\diff{r_v}/\diff\ell=r_v/\ell$. We can use this to
differentiate any function in the form $G(\rho)=\sum_v{g(r_v)}$,
leading to
$\diff{G}/\diff\rho=-(1/D\rho)\sum_v{r_v(\diff{g(r_v)}/\diff{r_v})}$.
From the derivatives of $U_{int}$ in Eq.~(\ref{novoa}), we obtain
\begin{equation}\label{eq:fp}
a(\rho)=\frac{1}{2D^2\rho}\sum_vr_v \left[(D-1)f(r_v) - r_v f'(r_v) \right].
\end{equation}
From Eq.~(\ref{eq:fp}), the internal force over a particle is obtained
from the surrounding concentration of particles $\rho$ and local
gradient concentration $\nabla\rho$. The sum in Eq.~(\ref{eq:fp}) goes
over all the vertices of a homogeneous lattice of concentration
$\rho$, and includes terms on the magnitude of interaction force
$f=-\diff{V}/\diff{r}$ as well as its derivative
$f'=\diff{f}/\diff{r}$. For most interaction potentials, this sum can not be analytically determined. However, knowing
the homogeneous lattice and interaction, it is a simple task to use
Eq.~(\ref{eq:fp}) to obtain $a(\rho)$ numerically for any
concentration.

\begin{figure}[t]
\centerline{\psfig{file=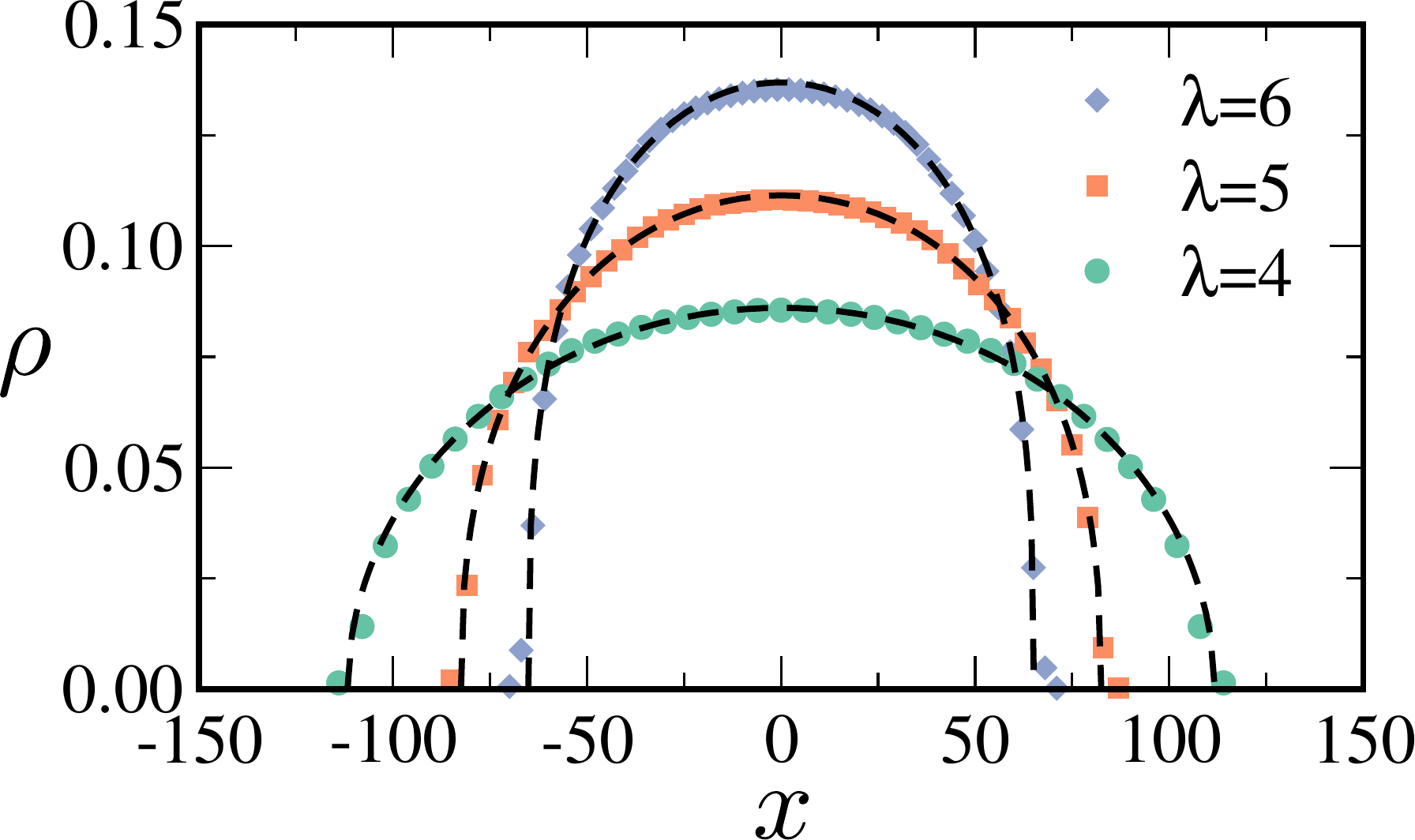, width=0.95\linewidth}}
\caption{Density profiles at the stationary state obtained from
  simulations (symbols). We consider two-dimensional systems of
  $N=900$ particles interacting through a power-law repulsive
  potential $V_\lambda(r)=\varepsilon\sigma^\lambda{r^{-\lambda}}$. In
  the $x$-direction, the particles are confined by a quadratic
  potential $U_{ext}(x)=kx^2/2$, with $k=10^{-5}
  \varepsilon\sigma^{-2}$. In the $y$-direction the simulation cell
  has a dimension $L_y=60\sigma$, with periodic boundary conditions.
  We present results for simulations of this system considering three
  different values of $\lambda$. The dashed lines represent the
  results of the continuum model via Eq.~(\ref{eq:solution_lambda}). }
\label{fig:perfis_456}
\end{figure}

In the case of a two-dimensional system of particles interacting
through a power law,
$V_\lambda(r)=\varepsilon\sigma^\lambda{r^{-\lambda}}$, for all
concentrations the homogeneous system rests in a triangular lattice
with lattice parameter $\ell=2^{1/2}/(3^{1/4}\rho^{1/2})$, leading to
\begin{equation}\label{c_lamb}
a_\lambda(\rho)=\rho^{\frac{\lambda}{2}-1}\left[\frac{3^{\frac{\lambda}{4}}(2+\lambda)\lambda\varepsilon\sigma^\lambda}{2^{\frac{\lambda}{2}+3}}\sum_v \left(\frac{\ell}{r_v}\right)^{\lambda}\right].
\end{equation}
Note that $\sum_v\ell/r_v$ is independent of $\rho$, and
Eq.~(\ref{c_lamb}) is consistent with our qualitative
prediction,~Eq.~(\ref{eq:a_errado}).

We now test our approach by comparing its predictions with numerical
simulations. In our first test we let $N$ particles interact in a
two-dimensional system, confined in the $x$-direction by an external
potential $U_{ext}(x)$, and with periodic boundary conditions in the
$y$-direction. We stop our simulation when the system reach
mechanical equilibrium, that is
\begin{equation}\label{eq:equilibrium}
F_{int}+F_{ext}=-a(\rho)\frac{\diff\rho}{\diff x}-\frac{\diff U_{ext}(x)}{\diff x}=0.
\end{equation}
One can then determine $\rho(x)$ by solving Eq.~(\ref{eq:equilibrium})
with the condition $L_y\int\rho\diff x=N$, where $L_y$ is the
transverse dimension of the simulation cell. For the case of a
power-law interaction,
$V_\lambda(r)=\varepsilon\sigma^\lambda{r}^{-\lambda}$,
$a_\lambda(\rho)$ can be obtained from Eq.~(\ref{c_lamb}), and the
solution of Eq.~(\ref{eq:equilibrium}) is
\begin{equation}\label{eq:solution_lambda}
\rho(x)=\left[{\rho_o}^\frac{\lambda}{2} - \frac{\lambda}{2C_\lambda}U_{ext}(x) \right]^{\frac{2}{\lambda}},
\end{equation}
where we define the parameter independent on $\rho$,
$C_\lambda=a_\lambda(\rho)\rho^{1-\lambda/2}$. 
Figure~\ref{fig:perfis_456} shows that Eq.~(\ref{eq:solution_lambda})
follows closely the results from numerical simulations for different
values of $\lambda$.

For systems of particles interacting through power-law potentials, the
density of energy is a power of the concentration,
$\rho~U_{int}(\rho)\sim\rho^{1+\lambda/D}$. In this case, considering
$q=1-\lambda/D$, the internal energy density, minimized by the
overdamped dynamics has a correspondence with the density of entropy
in the framework of the generalized Tsallis thermostatistics,
$s_q(\rho)=(\rho^{2-q}-\rho)/(q-1)$ \cite{tsallis1,*tsallis2}, that
should therefore be maximized. In fact, taking the same steps followed
in \cite{george}, and first introduced in \cite{curadotsallis}, it is
possible to show that, as long as $a(\rho)\sim\rho^{(\lambda/n)-1}$,
as in Eq.~(\ref{c_lamb}), Eq.~(\ref{eq:continuity}) will drive the
system towards an equilibrium state described by Tsallis distribution
$P_q(\mathbf{r})\sim[1-(1-q)\beta{U_{ext}(\mathbf{r})}]^{\frac{1}{1-q}}$
\cite{tsallis2014,soares1,soares2}, thus generalizing the previous
result of \cite{george,nobre2,*curado} for any $q\leq 0$.

\begin{figure}[t]
\centerline{\psfig{file=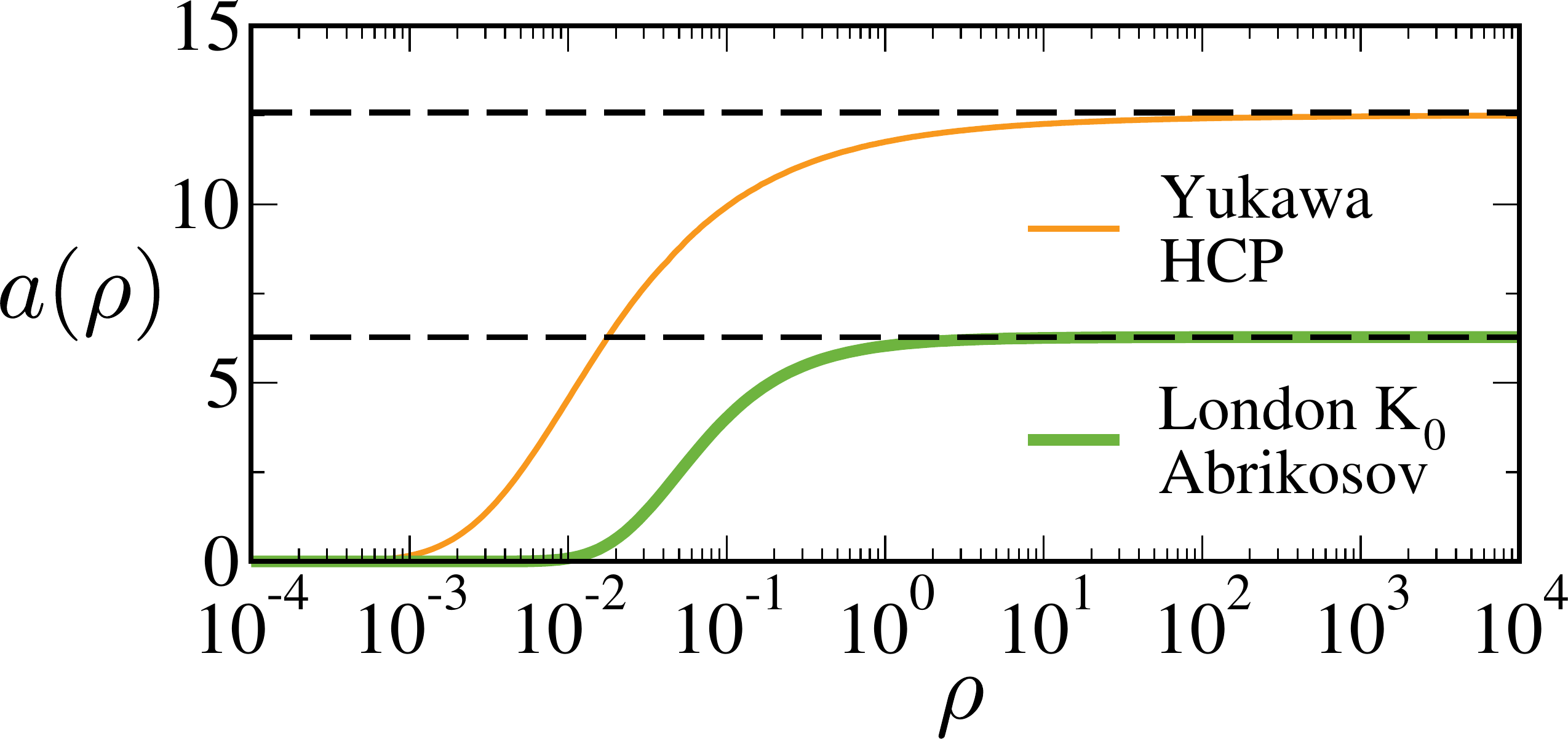, width=0.95\linewidth}}
\caption{The function $a(\rho)=|\mathbf{f}_{int}|/|\nabla\rho|$ as determined
  through Eq.~(\ref{eq:fp}). Two systems are investigated, namely,
  complex plasma particles interacting through the Yukawa potential,
  $V_Y=\varepsilon\sigma\exp(-r/\sigma)/r$, and forming Hexagonal
  Close-Packed (HCP) lattices in three dimensions; superconducting
  vortices interacting through the London potential,
  $V_S=\varepsilon{\mbox{K}_0(r/\sigma)}$, and forming triangular
  (Abrikosov) lattices. As indicated by the dashed lines, for high
  enough concentrations, $\rho>\sigma^{-D}$, the curves converge to
  the value given by Eq.~(\ref{eq:a_zapperi}), $a=4\pi\varepsilon\sigma^3$
  and $a=2\pi\varepsilon\sigma^2$, for the three and two-dimensional
  systems, respectively.}
\label{fig:arho}
\end{figure}

\begin{figure}[t]
\centerline{\psfig{file=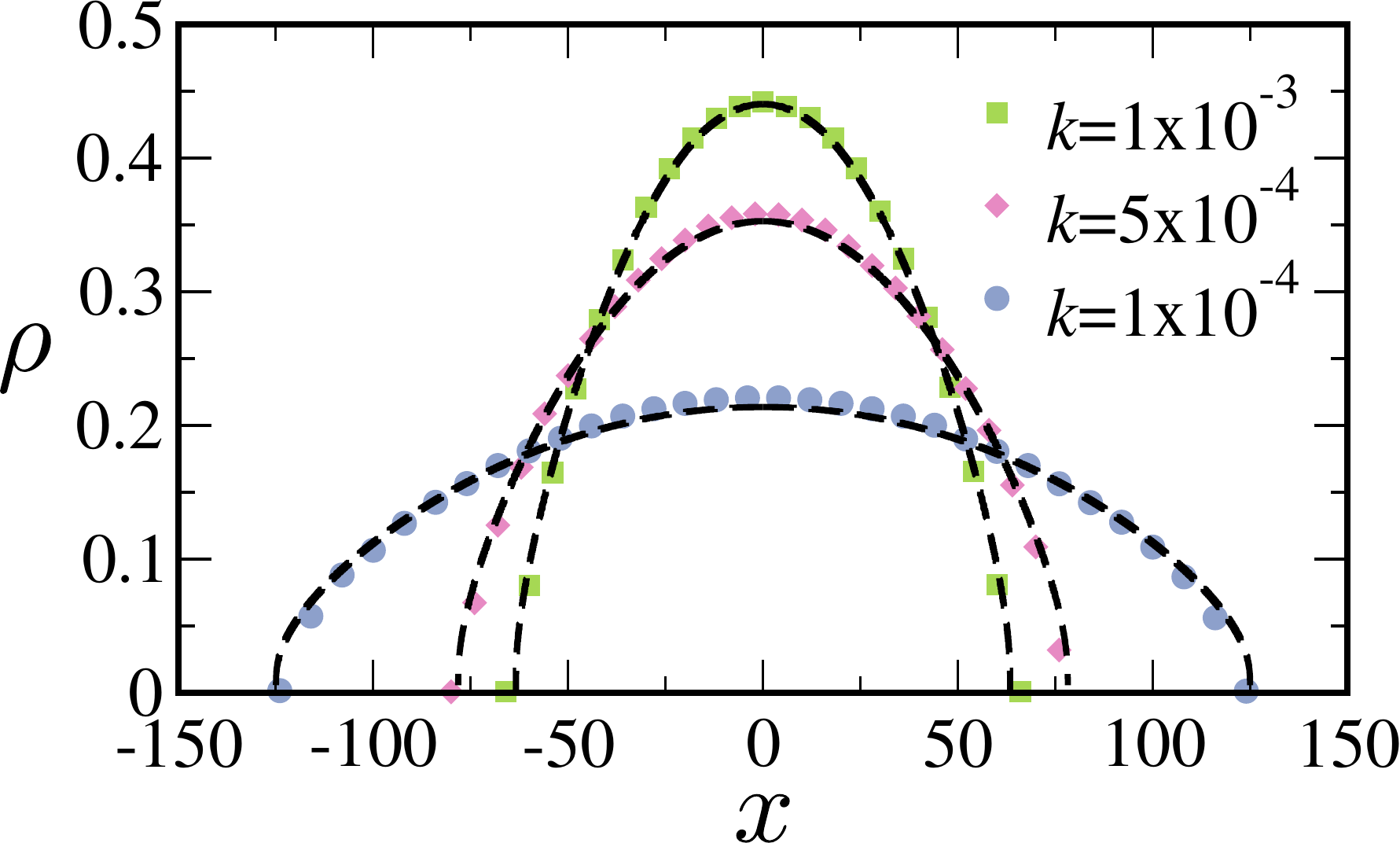, width=0.95\linewidth}}
\caption{Density profile at the stationary state obtained from
  simulations (symbols) for the London potential ($N=800$,
  $L_y=20 \sigma$) for three different values of $k$. The dashed
  curves represent the predictions of the continuum model. }
\label{fig:densgeorge}
\end{figure}

Next, we show applications of our approach to models of two physical
systems, namely, superconducting vortices and complex plasma. In
type-II superconductors the magnetic flux is confined to small regions
of the superconductor, each region being a vortex of super-currents
carrying one quanta of magnetic flux. Since the vortices cross the
sample system in the direction of the applied magnetic field, this
corresponds to a quasi two-dimensional system. Interacting through the
so-called London
potential~\cite{gennes,ref10_1,ref10_2,ref11_1,ref11_2},
$V_S(r)=\varepsilon{\mbox{K}_0}\left(\frac{r}{\sigma}\right)$,
vortices dissipate energy when moving. The least-energy state of this
system is a triangular lattice, also called Abrikosov
lattice~\cite{abrikosov}. Figure~\ref{fig:arho} shows the function
$a(\rho)$, computed through Eq.~(\ref{eq:fp}), for this system. For
larger concentrations ($\rho>\sigma^{-2}$) one sees that the function
saturates at the value predicted by Eq.~(\ref{eq:a_zapperi}), namely,
$a=2\pi\varepsilon\sigma^2$. However, for smaller concentrations the
particle-particle correlations become relevant and $a(\rho)$ goes to
zero as $\rho$ decreases. We test this result by simulating systems
confined in one direction by a potential $U(x)=kx^2/2$, and with
periodic boundary condition in the other
direction. Figure~\ref{fig:densgeorge} displays the solutions of
Eq.~(\ref{eq:equilibrium}) with the function $a(\rho)$ computed for
this particular system, and shows that they follow closely the results
from numerical simulations.

The other physical system we investigate is a complex plasma, namely, a
colloidal mixture of microscopic charged particles suspended in an
ionized gas. The free charges in the gas screen the Coulomb
interaction, and the Yukawa potential,
$V_Y=\varepsilon\sigma\exp(-r/\sigma)/r$, is a good model for the
repulsion between these colloidal
particles~\cite{peeters}. Considering that the ground state of this
system is an Hexagonal Close-Packed HCP lattice~\cite{chu,hcp}, we can
compute the function $a(\rho)$ as also shown in
Fig.~\ref{fig:arho}. Similarly to the case of type-II superconducting
vortices, for $\rho>\sigma^{-3}$, the function converges to
$a=4\pi\varepsilon\sigma^3$, in agreement with
Eq.~(\ref{eq:a_zapperi}), while going to zero as the concentration
$\rho$ decreases. To test this result, we simulate a three-dimensional
system of such particles under the action of an external potential
$U_{ext}(\mathbf{r})=kr^2/2$ confining the particles in all three
dimensions. As before, Eq.~(\ref{eq:equilibrium}) is numerically
solved, but the normalization must be imposed by
$4\pi\int\rho(r)r^2\diff{r}=N$.

\begin{figure}[t]
\centerline{\psfig{file=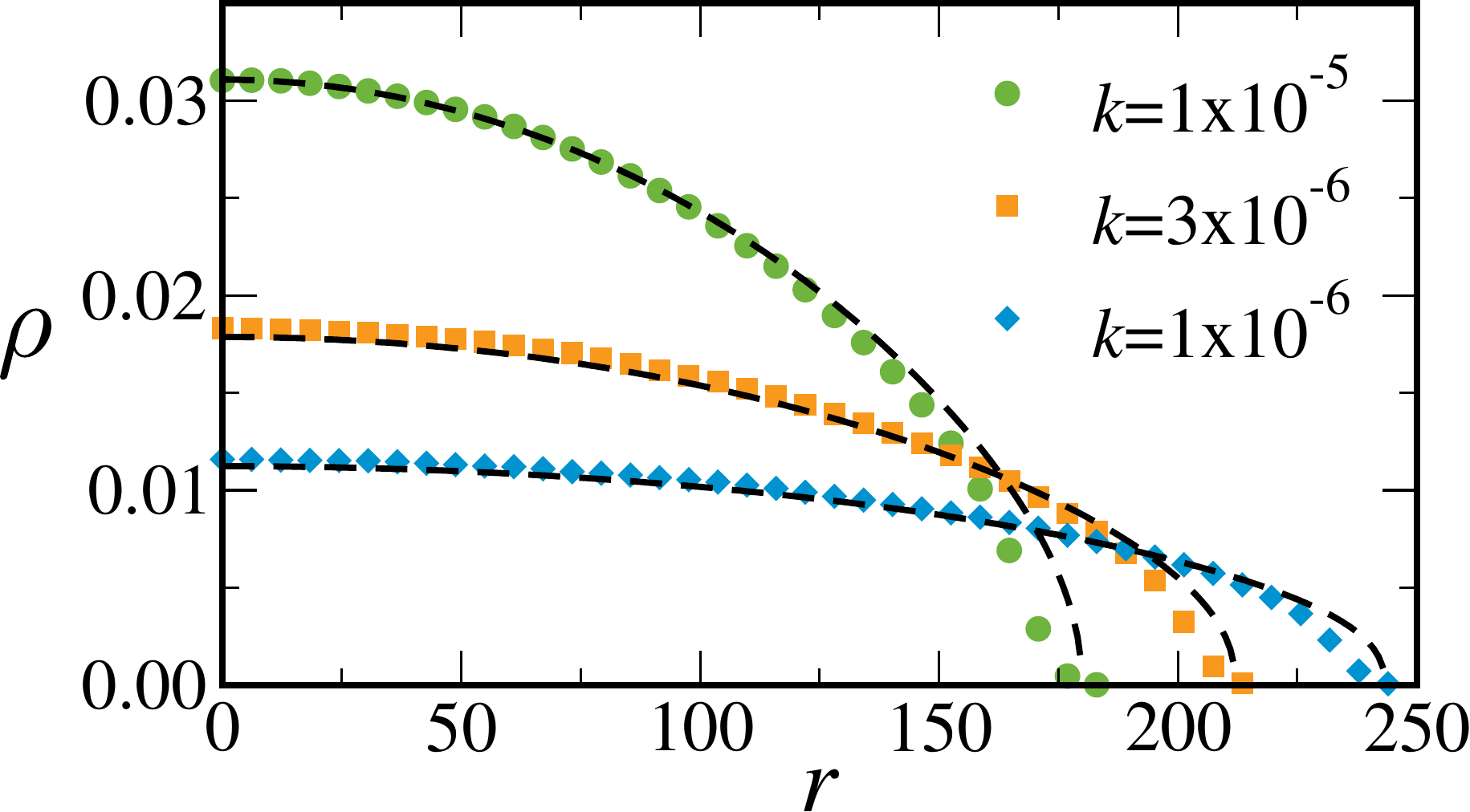, width=0.95\linewidth}}
\caption{Density profiles at the stationary state obtained from
  simulations (symbols). We consider three-dimensional systems of
  $N=400~000$ particles interacting through the Yukawa potential,
  $V_Y=\varepsilon\sigma\exp{(-r/\sigma)}/r$. The particles are confined
  by a quadratic potential, $U_{ext}(r)=kr^2/2$, with results for
  three different values of $k$. The dashed lines represent the
  predictions of the continuum model. }
\label{fig:teste3d}
\end{figure}

The comparisons between numerical simulations and the theoretical
predictions for the case of complex plasmas are shown in
Fig.~\ref{fig:teste3d}. For the bulk of the system we observe good
agreement between theory and simulation, but at the edge of the
density profile there is a notable deviation. To understand this
deviation, note that we assumed in our approach that the particle
concentration gradient varies slowly within the effective range of the
interaction potential. Since there is no negative concentration, this
assumption fails when the distance to the edge of the profile
$\rho/|\nabla\rho|$ is smaller than the characteristic interaction
length $\sigma$. Thus, at the edge of the profile a continuum
description should demand higher orders of approximation. However,
this effect becomes negligible in systems where the density profile is
not subjected to a strong confinement \cite{andradereply}.

In summary, we introduced a general approach to build continuum models
for systems of repulsive particles in dissipative media. For the two
physical systems investigated here, namely, superconducting vortices
and complex plasma, we show how the function $a(\rho)$, relating the
gradient of concentration to the force, converges at high
concentrations ($\rho>\sigma^{-D}$) to the value predicted by
Eq.~(\ref{eq:a_zapperi}). Therefore, the assumption of a constant
ratio force/gradient represents a good approximation for several cases
of interest, specially for highly concentrated
systems~\cite{zapperi2001flux,prb_moreira,*petrucio}. However, higher
concentrations may be a practical impossibility, specially in the case
of superconducting vortices, where the critical field imposes a
constraint in the maximum concentration of vortices~\cite{gennes}. For
systems of low concentration, or for interactions such as power-laws,
where Eq.~(\ref{eq:a_zapperi}) diverges, it is necessary to account
for the variation of the ratio $a(\rho)$ with concentration, as
proposed here.

\begin{acknowledgments}
  We thank the Brazilian agencies CNPq, CAPES, FUNCAP, and the
  National Institute of Science and Technology for Complex Systems
  (INCT-SC) in Brazil for financial support.
\end{acknowledgments}

\bibliographystyle{apsrev4-1.bst}
\bibliography{biblio}

\end{document}